**Full citation**: Felizardo, K.R., Salleh, N., Martins, R.M., Mendes, E., MacDonell, S.G., & Maldonado, J.C. (2011) Using visual text mining to support the study selection activity in systematic literature reviews, in *Proceedings of the 5th International Symposium on Empirical Software Engineering and Measurement (ESEM2011)*. Banff, Canada, IEEE Computer Society Press, pp.26-35. doi: 10.1109/ESEM.2011.16

# Using Visual Text Mining to Support the Study Selection Activity in Systematic Literature Reviews

Katia Romero Felizardo

*Department of Computer Science*
*University of São Paulo – USP*
*São Carlos, Brazil*
katiarf@icmc.usp.br

Norsaremah Salleh

*Depart. of Computer Science*
*International Islamic University*
*Malaysia*
norsaremah@iium.edu.my

Rafael M. Martins

*Depart. of Computer Science*
*University of São Paulo*
*Brazil*
rafaelmm@icmc.usp.br

Emília Mendes

*Department of Computer Science*
*University of Auckland*
*Auckland, New Zealand*
e.mendes@auckland.ac.nz

Stephen G. MacDonell

*SERL, Comp. & Math. Sciences*
*AUT University*
*Auckland, New Zealand*
stephen.macdonell@aut.ac.nz

José Carlos Maldonado

*Department of Computer Science*
*University of São Paulo – USP*
*São Carlos, Brazil*
jcmaldon@icmc.usp.br

**Abstract**

***Background***: *A systematic literature review (SLR) is a methodology used to aggregate all relevant existing evidence to answer a research question of interest. Although crucial, the process used to select primary studies can be arduous, time consuming, and must often be conducted manually.* ***Objective***: *We propose a novel approach, known as 'Systematic Literature Review based on Visual Text Mining' or simply SLR-VTM, to support the primary study selection activity using visual text mining (VTM) techniques.* ***Method***: *We conducted a case study to compare the performance and effectiveness of four doctoral students in selecting primary studies manually and using the SLR-VTM approach. To enable the comparison, we also developed a VTM tool that implemented our approach. We hypothesized that students using SLR-VTM would present improved selection performance and effectiveness.* ***Results***: *Our results show that incorporating VTM in the SLR study selection activity reduced the time spent in this activity and also increased the number of studies correctly included.* ***Conclusions***: *Our pilot case study presents promising results suggesting that the use of VTM may indeed be beneficial during the study selection activity when performing an SLR.*

## I. INTRODUCTION

The systematic literature review (SLR) is recognized as one of the key components of the Evidence-Based Software Engineering (EBSE) paradigm [1]. It provides reliable means and established methods to conduct a comprehensive and robust literature review based on three clearly defined phases [1]: (i) planning the review; (ii) conducting the review; and (iii) reporting the review. During the planning phase, the need for a review is identified and the review protocol is developed. The activities during the second phase include the identification of relevant research, selection of primary studies based on the inclusion and exclusion criteria, assessment of study quality, data extraction and data synthesis. Finally, the third phase includes dissemination or reporting of the SLR's results to interested parties including researchers and practitioners [2].

The substantial growth in the number of SLRs being undertaken in the area of software engineering (SE) lends weight to the importance of carrying out SLR activities effectively and efficiently [3]. Due to the comprehensive and rigorous nature of the work required when performing an SLR, its undertaking may be difficult and time consuming principally because some of the activities are conducted manually. In particular, the selection of primary studies can be arduous when an SLR involves a large volume of possibly relevant studies; consequently, it can be challenging to read, evaluate, and synthesize the state of the art of a particular topic of interest. It would seem advantageous to have a range of tools or techniques that could support the SLR activities, in particular the study selection activity. We contend here that the exploration and analysis of a large set of primary studies can be supported by the technique of Visual Text Mining (VTM) [4].

Text Mining is a well-established practice commonly used to extract patterns and non-trivial knowledge from unstructured documents or textual documents written in a natural language [5]. As an extension of this idea, Visual Text Mining (VTM) is the association of mining algorithms and information visualization techniques that support visualization and interactive data exploration [6]. In recent years, there has been an increasing interest in the use of visualization techniques as supporting tools for SLRs [7, 8, 9, 10]. This interest is motivated by the fact that humans present strong visual processing abilities; therefore visual-based techniques make use of these abilities, by using the human system to support knowledge discovery [4] i.e., in the SLR context, to discover the relevant primary studies.

The aim of this paper is to propose and evaluate an approach based on VTM, the SLR-VTM (Systematic Literature Review based on Visual Text Mining) to support the selection of primary studies in an SLR, offering different perspectives to those considered in prior work i.e., visualizations based on the content of the studies and based on their citation relationship. The specific contributions of our work to the body of knowledge in this field are the following: (i) a new VTM approach containing different visualizations and strategies to assist the selection of primary studies; (ii) an automated tool to support the selection activity in SLRs; and (iii) empirical evidence regarding the effects of the use of VTM in an EBSE context.

The remainder of this paper is organized as follows. In Section 2, background information is provided regarding related work on both the SLR process and VTM; the motivation of this research is also elaborated. Section 3 presents our proposed approach, as well as a supporting tool, named Revis. In Section 4 we detail the research methodology employed in our pilot case study, followed by the detailing of results, lessons learned and limitations of this work in Section 5. Finally, Section 6 concludes our work.

## II. BACKGROUND AND RELATED WORK

The bibliographical or informal literature reviews frequently seen in the literature do not use a systematic approach; hence one cannot rule out that the choice of studies and the conclusions drawn could be biased, thus providing readers with a distorted view about the state of knowledge regarding the area at the focus of the review. Conversely, an SLR uses a systematic process to identify, assess and interpret all available research evidence aiming at providing reliable answers to a particular research question [11], [3]. An SLR is considered to be a sound methodology that can be used to identify publications related to a specific subject via a predefined search strategy aimed at minimizing bias [1]. Considering its growing relevance in the field of software engineering (SE), this methodology has been increasingly applied by researchers and practitioners in various topics of interest [2], [12]. In addition, due to the increasing number of primary studies in SE, it would be advantageous to have an efficient mechanism to summarize and provide an overview about an area or topic of interest in this field [13].

Knowledge Discovery in Databases (KDD) has been applied in other fields to extract high-level, potentially useful knowledge from low-level data [4]. Data Mining (DM), which is one of the components of the KDD process, has been used to extract patterns or models from data. In order to support the KDD process, a visualization technique can be combined with DM, resulting in a Visual Data Mining process (VDM) [4], [6]. In VDM, a visualization technique supports user interaction with the mining algorithm, facilitating productive and effective discovery methods [4]. By extension, Visual Text Mining (VTM) refers to VDM applied in text or to a collection of documents [6]. VTM combines text mining algorithms with interactive visualizations. As such, we contend that it can provide useful support to users who need to make sense of a collection of primary studies, helping them to decide which studies to include in an SLR.

Several studies have investigated the potential benefits of visualization in supporting the conduct of an SLR. El Emam et al. [7] investigated the use of Electronic Data Capture (EDC) tools to provide automated support for data collection and query resolution, among other features, for clinical trials during an SLR process. Despite their focus being on the use of electronic data capture tools, the study selection activity is still manually conducted. Ananiadou et al. [9] employed text mining tools to support three different activities of the SLR process: (i) search, (ii) study selection activity – using document classification and document clustering techniques – and (iii) syntheses of the data; however their focus is in the social sciences field. In addition, two previous studies [8, 10] have specifically investigated the use of VTM within the context of EBSE. Felizardo et al. [8] employed VTM to support categorization and classification of studies when carrying out systematic mapping studies; and Malheiros et al. [10] investigated the use of VTM techniques to help with SLRs, using a feasibility study. They compared the performance of reviewers in selecting primary studies using two different methods: (i) reading abstracts; and (ii) using a VTM approach. Their results were promising, showing that the use of VTM can both reduce the time required and improve the effectiveness of the selection of primary studies. However, a time limit was imposed on the conduct of their experiment. In addition, none of these previous studies has investigated the use of meta-data analysis, for example, citation network, to assist the selection activity.

Similar to Malheiros et al.'s work [10], the approach presented herein also makes use of VTM to support the study selection activity in the process of SLR. However, our work differs from that of Malheiros et al. by supporting the selection of primary studies using content-based analysis of documents (document map), and meta-data analysis, via representations such as edge bundles and citation networks. In their study, Malheiros et al. used the document map as a single visualization technique to support the selection activity. The document map used as its basis the content of the papers (i.e., titles, abstracts, keywords) whereas in our case, the other two visualization techniques used citations among studies and shared references between studies to indicate citation relationships. Such information would appear to be useful in supporting the decision process when selecting papers.

## III. SYSTEMATIC LITERATURE REVIEW BASED ON VISUAL TEXT MINING (SLR-VTM)

The Systematic Literature Review based on Visual Text Mining (SLR-VTM) is an approach to support primary studies' selection during the SLR process. As illustrated in Figure 1, the approach comprises four stages: (i) planning; (ii) search process; (iii) visualization; and (iv) VTM selection. The first stage is carried out according to procedures defined by Kitchenham and Charters [1]. During Stage 1 (Planning), the SLR protocol is defined. The protocol contains, for example, the research question(s), source search methods, inclusion and exclusion criteria and the primary studies' selection process, among others. In Stage 2 (Search process), primary studies are identified using each of the source searches previously documented in the protocol. Stage 3 (Visualization) involves the generation of visual representations of the primary studies obtained in Stage 2. In Stage 4 (VTM Selection), the relevant primary studies are selected (or discarded) applying the inclusion and exclusion criteria. We have investigated the application of VTM to Stages 3 and 4, discussed in more detail next.

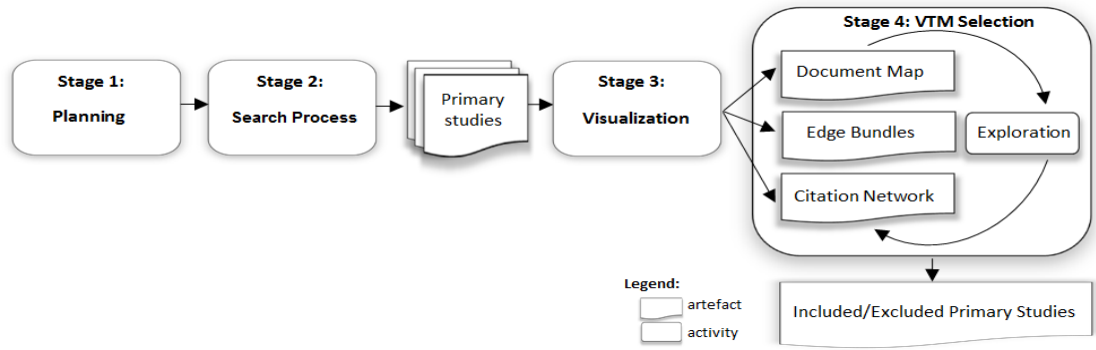

**Figure 1**. Systematic Literature Review based on VTM.

### A. Visualization Stage

In this stage three visual representations of the primary studies previously selected are generated: (i) a document map; (ii) edge bundles and (iii) a citation network. Each is described as follows:

- A **document map** (see Figure 2a) is a 2D visual representation of the primary studies, and enables users to investigate content and similarity relationships between these studies. The process to create a document map involves the conversion of all primary studies selected (title, abstract and keywords) into multi-dimensional vectors. The dimensionality is based on all the terms extracted from the primary studies. The dimensionality of these vectors can be reduced eliminating stopwords (i.e., minimally representative terms, such as prepositions, articles and conjunctions), applying stemming (i.e., converting terms to their radical; for instance, 'testing' and 'tester' are both reduced to 'test') and using projection techniques [14]. In short, projection techniques map each primary study to a graphical element represented by a circle; circles are then placed on the 2D layout in a way that reflects similarity relationships. Thus, similar documents are placed close to one another and dissimilar documents are positioned far apart. The document map also shows regions that group primary studies based on their similarity. Similarity is calculated using the cosine similarity measure, often used to compare documents in text mining. It is a measure of similarity between two vectors of n dimensions by finding the cosine of the angle between them, which ranges between 0 (no similarity) and 1 (completely similar) [15]. More details about the process to create the document map can be found in Felizardo et al. [8].

- The **edge bundle** was defined by Holten [16] as a hierarchical tree visualization technique that shows both nodes and node-links (relationships between nodes) at the same time. In our case, the nodes (small circles, see Figure 2b) are the primary studies and the node-links (blue lines) are the citations between them. In order to create the hierarchical tree we used the HiPP (Hierarchical Point Placement) strategy [17] and the node-links were coloured to represent the direction of the citation: the citing paper is at the light blue end of the link and the cited paper at the dark blue end.

- Finally, the **citation network** shows the primary studies (central point – circle) with their cited references (circles around the central point, connected by edges). Through this depiction it is possible to see citations between the primary studies with their own references and also citations between primary studies and references of other primary studies (references shared – see Figure 2c). The citation network visualization uses a force-based algorithm [18] to position the points on the layout. This means that studies attract or repel one another depending on how strong their connections (references to each other) are. Primary studies that do not share references (isolated primary studies) are disconnected from the other studies in the network.

## B. VTM Selection Stage

During this stage the primary studies' selection activity takes place. We propose that the VTM approach can support the selection activity using three different visualization methods; their respective strategies of exploration are detailed next.

- **Document Map:** There are three VTM techniques (see Figures 3b, 3c and 3d) that can be applied to a document map (see Figure 3a):

  o **Clusters and Topics:** One strategy to classify primary studies is to identify the regions (clusters) of documents with similar content in terms of their titles, abstracts and keywords. Using this technique, clusters are created automatically followed by the formation of their associated topics. These topics are labels that represent the content of the documents contained in the clusters. In order to efficiently include groups of primary studies, a user can concentrate their reading on documents that belong to the clusters labeled with topics that most closely match the SLR's research questions. Similarly, in order to exclude studies, a user can review (perhaps less thoroughly) the documents belonging to clusters labeled with topics that do not match their SLR's research questions. Figure 3b shows a document map after the clusters and topics technique was applied. The colour of each point represents the cluster it belongs to and topics appear inside boxes.

  o **Expression Occurrence:** This technique changes the colour of each point in the document map in order to represent the frequency of occurrence of specific user-defined expressions in the primary studies. In this case, the colour scale varies from black (i.e., no occurrence) to white (i.e., many occurrences). A user can then prioritize their reading towards documents coloured in white (or closer to white) in order to consider whether these documents should be included in the SLR. Conversely, a user can read the documents coloured in black (or closer to this colour) to determine whether those documents should indeed be excluded from the SLR (assuming, of course, that the user-defined expression is relevant). Figure 3c illustrates the coloured document map using this technique, where the white point indicates the maximum occurrence of an expression.

  o **KNN Edges Connection (Neighborhood Relationship):** This technique connects primary studies with their neighbors to support study inclusion by association. That is, the closer the neighbors of an included study are the more likely they are to be also relevant to the SLR. Likewise, the neighbors of an excluded study are more likely to be irrelevant and so should not be included. Figure 3d shows the points connected to identify their neighborhood.

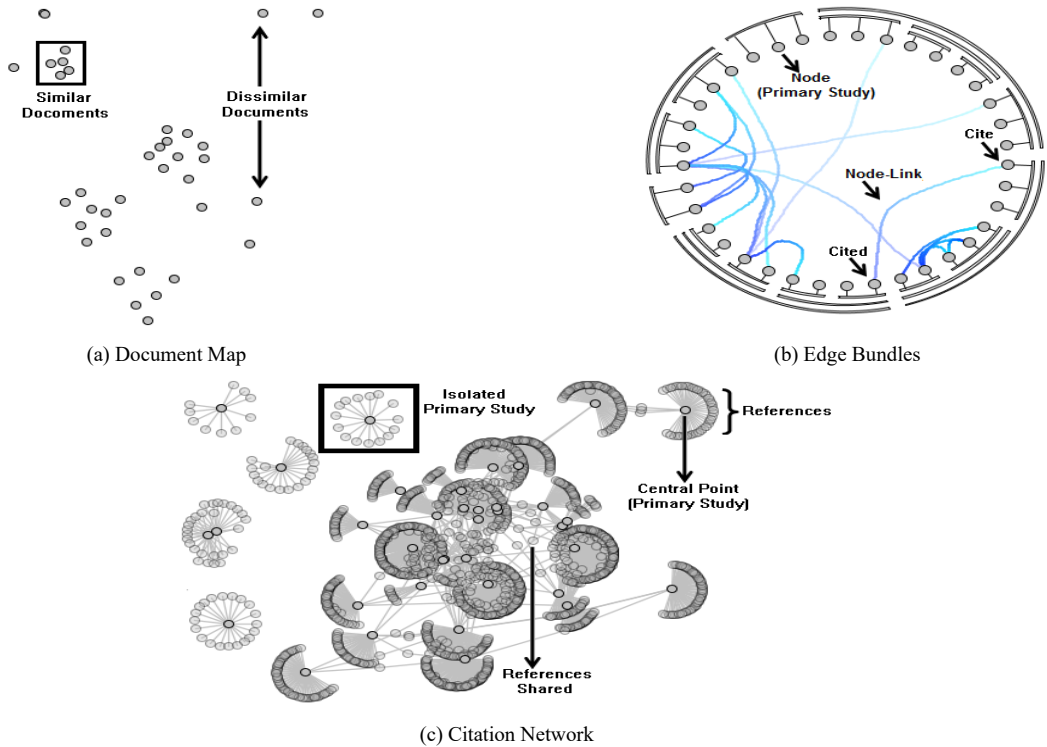

(a) Document Map

(b) Edge Bundles

(c) Citation Network

**Figure 2**. The three different views created in the Visualization Stage.

a. In general, visualization techniques employ colour in order to add extra information on a visual representation. We suggest the reading of a colour print version of this paper.

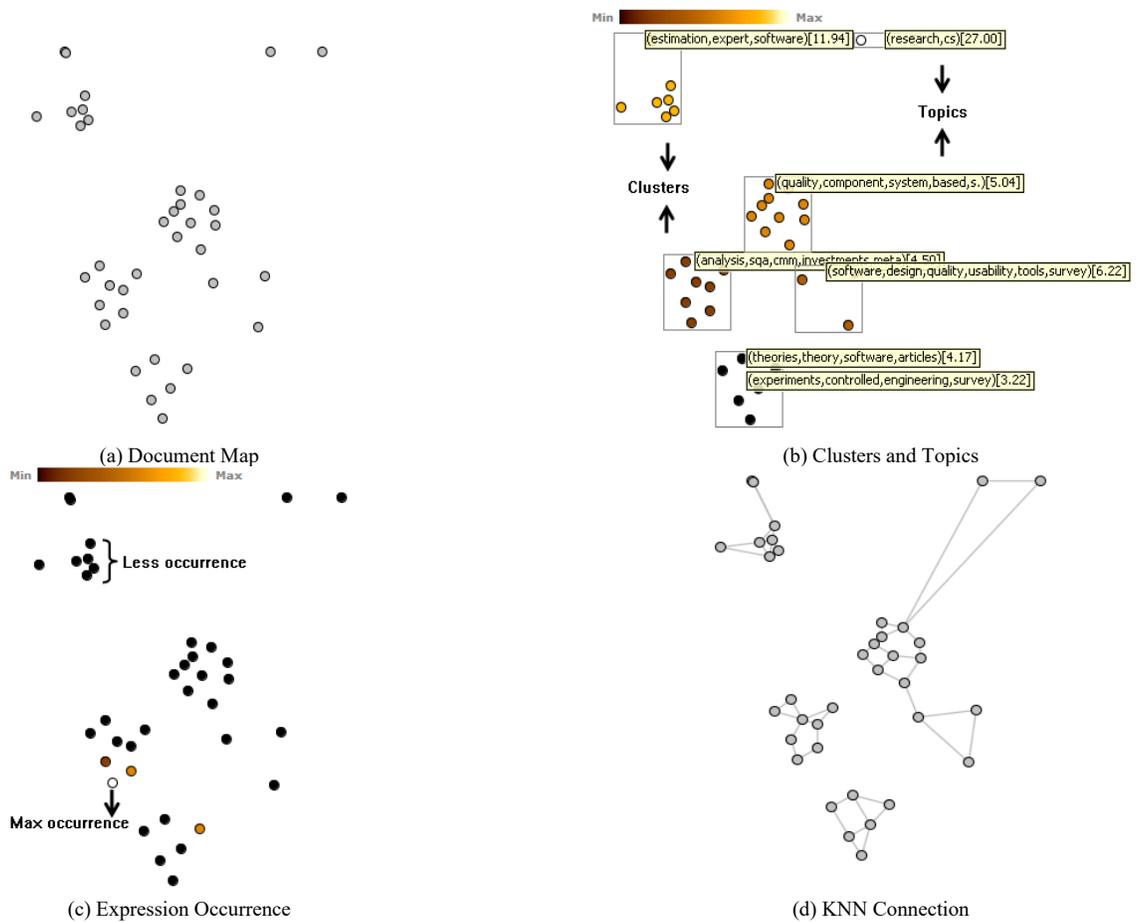

Figure 3. The three different VTM techniques (b-d) that can be applied to the document map (a).

- **Edge Bundles:** A relevant paper is usually cited by other papers. The edge bundles can be used to visualize the number of times that a paper is cited and papers that are cited many times are strong candidates to become primary studies to include in the SLR, or at least given due attention by the user. On the other hand, papers that are not cited, or cited very little, may be indicative of studies that should not be included in the SLR.

- **Citation Network:** This view offers important additional information beyond that related to the initial set of documents, in particular a primary studies' references and the connections between papers via the set of references that they share. Reference lists from relevant primary studies could be other sources of evidence to be searched [1]. Hence, papers that share references with a relevant paper could be more appropriate for inclusion in the SLR. On the other hand, primary studies that are not connected to any other studies (i.e., do not share citations or references), are more likely to be irrelevant documents in terms of the research question and may therefore be more readily excluded from the SLR.

The strategies just mentioned can be applied iteratively in order to adequately seek primary studies, as depicted in the various visual representations. The number of iterations is determined by the user and should continue until all primary studies have been considered. Moreover, the user can combine the techniques using **coordination**, which represents an interaction among the different views (i.e., document maps, edge bundles, and citation network). Using coordination, once a point or a group of documents in a view is selected, the corresponding point (or points) is then highlighted in the other views. Figure 4 illustrates the coordination of the three views. In Figure 4a, the document numbered as 1 is cited most frequently (see edge bundles view, Figure 4b) and it has the maximum occurrence of a specific and relevant expression in the context of their SLR question research (see document map view, Figure 4a). Hence, the user has additional information about document 1 that could support their decision on whether to include the study in the SLR. The citation network (see Figure 4c) shows that document 1 and its closest neighbors (i.e., documents 2 and 3) also share their references. Analysis of the edge bundles view (Figure 4b) shows that document 1 is cited by document 2.

Thus, if document 1 is considered appropriate to be included in the SLR, then it is likely that documents 2 and 3 are also to be included. Various other combinations can be used all aiming at supporting inclusion/exclusion decisions. Users should employ the views to obtain additional information in relation to the document set or to individual studies. Note that at any point a user can refer to the abstract/full text of any study, by double clicking on the relevant point in a visual representation. In the next section, we present our tool that can be used to automate Stages 3 and 4 of the SLR-VTM approach.

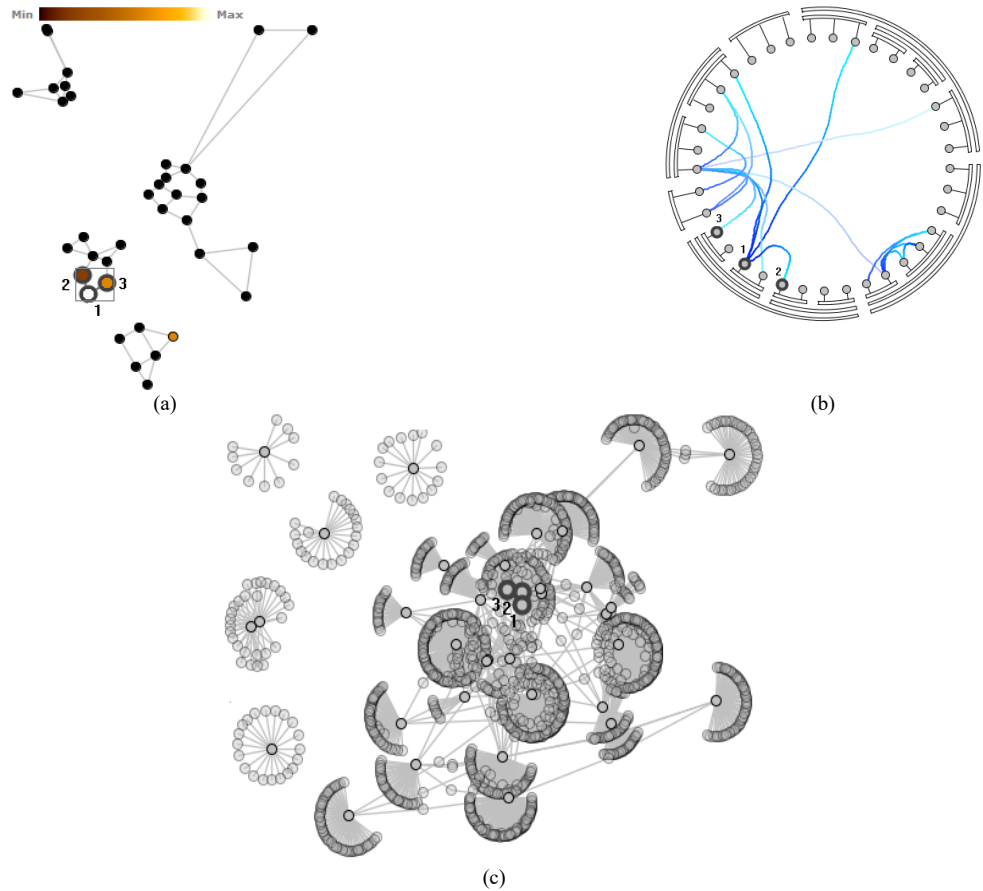

(a)

(b)

(c)

**Figure 4**. Coordination between the three visualizations.

## C. Support Tool: REVIS – Systematic Literature Review Supported by Visual Analytics

Revis – Systematic Literature Review Supported by Visual Analytics – is a flexible visualization tool that enables a user to leverage several VTM capabilities to explore a collection of documents (primary studies). Figure 5 shows the Revis tool's main window. Revis takes as input a set of primary studies selected during Stage 2 of the SLR-VTM approach. These studies are organized according to the *bibtex* format, which includes their title, abstract, keywords, and references. Revis then executes the activities performed during SLR-VTM Stage 3 and presents the document map, edge bundles and citation network for the document set.

The main VTM functionalities offered by Revis are described as follows: - it creates the views: document map, edge bundles and citation network; - it applies clustering algorithms in order to create clusters and their respective topics; - it allows changing of visual attributes (colour) of the points in the document map to represent the frequency of occurrence of an expression in the documents, or the status of the document (i.e., green circles represent the included studies and red the excluded); - it allows users to explore neighborhood relationships in the document map, with neighboring connections (called KNN connections) shown as edges between the points; - it supports coordination between different views, the document map, edge bundles and citation network; - it displays the content of a document (i.e., title, abstract and keywords) in a separate window when the user double clicks a node.

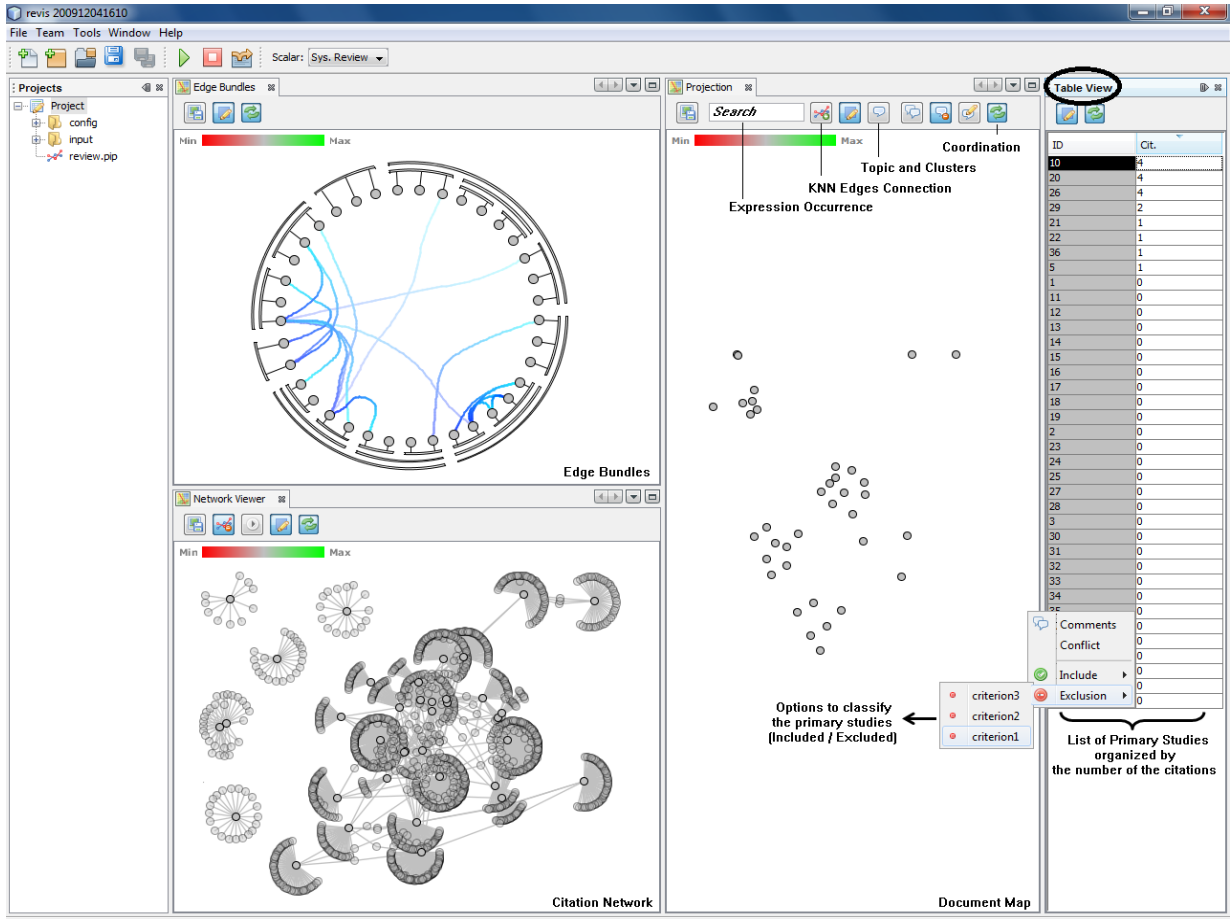

**Figure 5**. Main Window of Revis.

Excluding or including a document can be done easily by clicking the right-hand mouse button over the "Table View" pane and choosing one of the options shown in the pop-up window. This scenario is illustrated on the right pane of Figure 5. The points are coloured as red or green according to their exclusion or inclusion respectively. An example is illustrated in Figure 6. Note that, as described previously, those included documents that have similar content are positioned near to each other in the document map, illustrating that the neighbors of an included study are more likely to be relevant. In general they cited one another frequently and also share references in common. Figure 6 also shows that the neighbors of an excluded study are more likely to be irrelevant, and that most of the outliers (isolated studies, red point – see citation map (c)) are classified as excluded.

In the next section we present a pilot case study that empirically assesses the utility of the Revis tool and our proposed VTM techniques.

## IV. PILOT CASE STUDY: INVESTIGATING THE USE OF VTM TECHNIQUES TO SUPPORT STUDY SELECTION

In order to validate our SLR-VTM approach we conducted a pilot case study involving PhD students. In this paper, we argue that VTM techniques can support the SLR process' study selection activity. Previous findings reported by Malheiros et al. [10] show that the use of VTM reduces the time taken to perform the selection activity and also improves the number of papers correctly included or excluded in SLRs. However, their study used just one type of visualization (i.e., the document map). In our study, we applied this same visualization technique plus another two new visualization techniques (i.e., edge bundles and citation network). Hence, our empirical research questions (RQ) are:

- **RQ1:** Do VTM techniques (document map, edge bundles, and citation network) improve the performance (time taken) of the study selection activity in the SLR process?
- **RQ2:** Do VTM techniques improve the effectiveness (correctness of the

inclusion/exclusion) of the study selection activity in the SLR process?

The research objectives, the instruments and procedure used in this case study are detailed in the following subsections, followed by the results obtained.

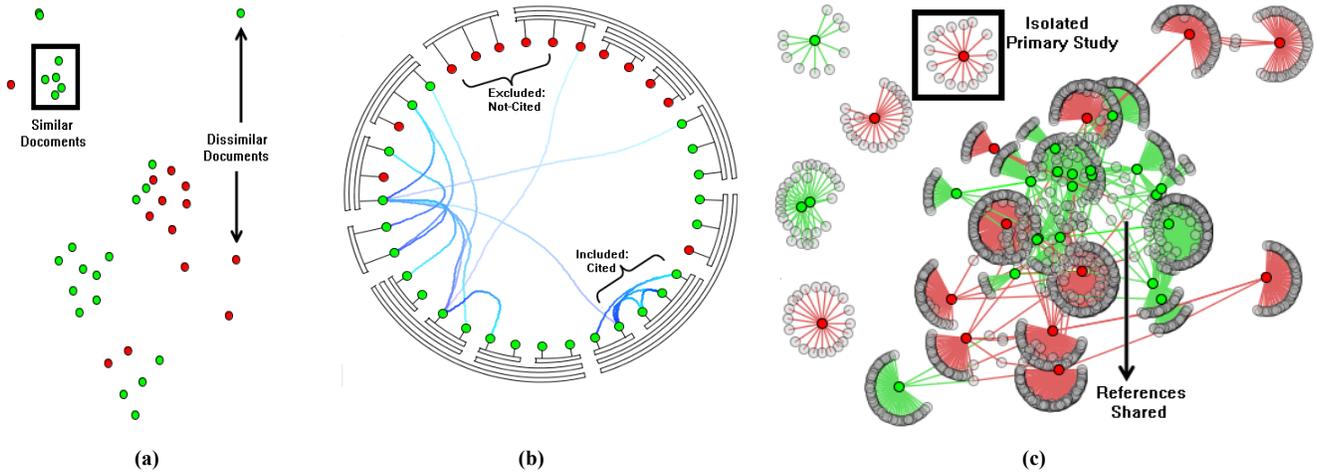

**Figure 6**. Views after completion of the selection activity, where included studies are coloured in green and excluded in red.

## A. Pilot Case Study Research Objectives

This case study aimed to investigate the use of VTM techniques to support the study selection activity of the SLR process. The objectives of our case study were outlined using the Goal/Question Metric template (GQM) [19] and the goal definition for the case study is the following:

- **Object of study:** VTM techniques.
- **Purpose:** To improve the performance and the effectiveness of the primary study selection activity in SLR process.
- **Focus:** To investigate the use of VTM techniques (using the Revis tool) during the study selection activity.
- **Perspective:** From the point of view of the researcher.
- **Context:** In the context of PhD student work.

We contend that the VTM techniques will positively influence the performance and effectiveness of PhD students undertaking the selection of primary studies. Therefore, we have investigated whether or not the use of VTM techniques affects the productivity of students who performed the selection of primary studies during an SLR. Four PhD students participated in this pilot case study, and all had prior experience in conducting SLRs.

## B. Instrumentation and Procedure

At the start of the case study, one of the authors provided all participants with an overview of the case study, which was organized in two stages: (i) training; and (ii) execution. Participants were split randomly into two groups, one to conduct the study selection activity manually (group 1) and another to conduct the study selection activity using the VTM approach (group 2). For training purposes, we used a small set of data (i.e., 20 primary studies) and a specific set of inclusion and exclusion criteria. Participants in group 1 classified the primary studies as included or excluded based on their reading of the abstracts. They were presented with the list of papers to be analyzed (the titles, abstracts, and keywords), the inclusion and exclusion criteria, and a table to summarize the decision whether to include or exclude a study.

Only participants in group 2 were trained on how to use the Revis tool so that they were familiar with the VTM techniques and the case study form. As part of the case study, they were given a description of the VTM techniques, the inclusion/exclusion criteria, and a table to summarize the decision whether to include or exclude a study. During the training stage, participants' doubts about the reading process, the tool, the VTM techniques or the form were clarified. To ensure that first impressions from the training would not interfere with the case study, we used a different and larger set of data (set 2) for the execution stage. During this stage, the primary studies selection activity was carried out using set 2. This set of studies came from an SLR conducted and double-checked by an expert in SLRs, who is a collaborator with our research group. We relied on their expert opinion to define the studies that should be included or excluded.

As in the training stage, group 1 read the abstracts and group 2 used the Revis tool and applied the VTM techniques. Participants were required to record the time they took to execute the selection activity. This allowed us to measure the time spent on the selection activity as an indicator of performance.

## C. Results

This section presents the results of our pilot case study in order to address our specific research questions (RQ1 and RQ2). A summary of results is presented in Table 1. Note that no statistical significance tests were used due to the very small sample employed.

To answer our first research question (RQ1), we measured participants' performance as the time they spent undertaking the execution stage i.e. the time spent by reviewers to make their decisions, which does not include the time required to prepare the data for the tool (see Table 1 – third column).

Our results showed that the time taken by students 1 and 2 to perform the study selection activity on the basis of reading the abstracts was 85 and 54 minutes respectively; and the time taken by students 3 and 4 to perform the same activity using the VTM approach was 30 and 58 minutes respectively. The performance of the students using the VTM appeared to be either equivalent to or better than that of the students using the manual method.

**TABLE I.** SUMMARY OF RESULTS.

| Approach | Student | Performance (Time spent) | Correctness | |
|---|---|---|---|---|
| | | | Included/ Excluded Correctly | Included/ Excluded Incorrectly |
| Reading Abstract (Group 1) | #1 | 85 min | 25 | 12 |
| | #2 | 54 min | 22 | 15 |
| VTM strategies (Group 2) | #3 | 30 min | 27 | 10 |
| | #4 | 58 min | 28 | 9 |

Table 1 (see fourth and fifth columns) shows the comparison between the VTM and manual reading approaches in terms of the number of studies correctly included/excluded. The number of studies included/excluded correctly using the manual reading approach was lower than if using the VTM approach. Students 1 and 2 (using the reading approach) correctly included/excluded 25 and 22 papers respectively, whereas students 3 and 4 (using the VTM approach) correctly included/excluded 27 and 28 papers, respectively. One important point about this result is that the students generally selected the same correct studies (i.e. the range of correctly identified studies did not vary widely). Considering the studies incorrectly judged, only student 2, who read the abstracts, had more false-negative decisions (i.e. relevant studies excluded – 10 studies) than false-positives (i.e. irrelevant studies included – 5 studies), where false-negative decisions are more serious and difficult to correct in an SLR than false-positives.

Our results therefore suggest that the use of the VTM approach can help to improve the performance of the study selection activity in the SLR when compared to a manual reading method. The results also show that the outcomes of the selection of the primary studies may be more reliable when using the VTM approach. Our evidence overall therefore suggests that the application of the VTM techniques is promising as, in this case, it maintained or improved both the performance and effectiveness of primary studies selection.

## V. DISCUSSION

In this section we discuss issues related to the results we obtained in our case study and the limitations or threats to the validity of our findings. Based on our results, we found that the application of the VTM techniques in the study selection activity has provided some benefits; in particular, it appears to accelerate the selection of primary studies and ameliorates the identification of relevant and/or irrelevant papers.

Keim [4] highlighted that visualizing and exploring information using VDM would help the user to deal with vast amounts of data, or in our case, large numbers of primary studies. This is because the visualization supports user interaction with the mining algorithm and directs it towards a suitable solution to a given task. Moreover, VDM can be used to enhance user interpretation of mining tasks [6]. The results of our case study lend support to their views since the VTM techniques (i.e., the application of VDM to a collection of texts) facilitates exploration, interpretation, and decision-making in regard to the inclusion or exclusion of primary studies.

The results obtained from our case study support previous findings reported by Malheiros et al. [10], and suggest that the VTM approach can speed up the selection process of primary studies, and that the precision/accuracy of the selection of relevant studies is at least as good as using the manual method. Although not part of the formal study, qualitative feedback from the case study participants indicated that the tool helped in minimizing the effort to select the primary studies. One of the participants from group 2 mentioned that the tool was very useful; it reduced the time spent to make a decision whether to include/exclude studies. Each of the three visualization techniques required similar mental effort to be understood and used. The group 2 participants also mentioned that by allowing users to explore different visual representations of primary studies, the SLR-VTM approach provides additional and complementary detailed information that is not readily available directly from reading the studies´ abstracts (e.g. similarity relationships, citations between primary studies).

The use of Revis to support the selection activity generally requires extra time to prepare and provide the information about the papers for input to the tool. The time taken depends on a number of factors, including: (i) the number of primary studies to be analyzed; (ii) the literature search used, for example, if the Web search engines provided by digital libraries do not utilize an automated search. Usually, an automated search retrieves results from search engines in a *bibtex* format, used by Revis; in other cases, if the studies are in any other format (e.g. PDF), it is necessary to convert them prior to the analysis; and (iii) the number of manual searches conducted.

### A. Threats to Validity

One of the potential threats to the internal validity of our study relates to the sample used in our pilot case study (four PhD student participants, who are, to some degree, influenced by their supervisors). In our view, a larger sample size of diverse SLR practitioners would help increase the reliability of the findings. As a first-cut assessment of the techniques, however, we believe our study is still useful. Typically, many SLRs involve a greater number of studies to be considered during the selection stage (more than 100). However, we chose to use in our case study an SLR published in the literature containing 37 primary studies. We made this choice on the assumption that adding too many studies in our case study could similarly influence the results because it might affect the motivation and performance of the participants in carrying out the assigned tasks. Despite the fact that our example case contains a rather small number of studies, the Revis and VTM approach suggested can be used in real SLRs, where a large number of candidate studies are considered – hundreds and even thousands. In fact, according to the VTM specialists, VTM tools work better with more articles [10].

One alternative explanation for the outcomes is that group 2 received training, which might have given them more confidence and understanding about the tasks being undertaken.

## VI. CONCLUSIONS AND FUTURE WORK

A systematic literature review commonly involves a large set of data to be analyzed and interpreted. In the SLR phases, the selection of primary studies is one of the most important activities that could impact the quality of the SLR's results. In light of this, the main contributions of this paper are as follows: (i) we developed a VTM approach and a tool (Revis) to support and partially automate the study selection activity of the SLR process; (ii) our evidence shows that the use of VTM to support the SLR process is promising (in terms of effort reduction and selection correctness). In conclusion, the work presented here extends Malheiros et al.'s work [10] in supporting the SLR-VTM approach. Both studies make use of VTM techniques to support study selection in the SLR process. We have proposed the use of new visualization techniques, in particular, edge bundles and a citation network that facilitate the exploration task. Additionally, the new visualizations can be used together with the document map through the coordination technique, also proposed here.

The results of our pilot case study indicate that the SLR-VTM approach was useful in accelerating the selection task. Furthermore it helped to increase the inclusion of relevant papers and the exclusion of irrelevant papers. In addition to larger replications of this work, we believe that interesting directions for future research are: (i) replicate the case study presented here involving more students, as well as with SLR experts; (ii) explore other types of VTM techniques and visualizations to further support the selection activity; (iii) test the individual effectiveness of the various representations, analyzing the trade-off between the number of visualizations and the incremental benefit to be gained from each one; and (iv) analyze whether Revis could provide support for validating inclusion/exclusion decisions.

## ACKNOWLEDGMENT

This research is supported by Brazilian funding agencies: CNPq (Processes n. 141972/2008-4; 201622/2009-2), CAPES and FAPESP (Process n. 2008/02024-6.